\documentclass{article}
\usepackage{dcase2024,amsmath,graphicx,url,times,booktabs,tabularx}
\usepackage{subcaption}
\usepackage{multirow}
\usepackage{siunitx}
\usepackage[ruled,noline]{algorithm2e}
\usepackage[inline]{enumitem}

\title{Frequency tracking features for data-efficient\\ deep siren identification}

\name{Stefano Damiano$^{1}\sthanks{This project has received funding from the European Union's Horizon 2020 research and innovation programme under the Marie Sk\text{\l}odowska-Curie grant agreement No. 956962, from KU Leuven internal funds C3/23/056, and from FWO Research Project G0A0424N. This paper reflects only the authors’ views and the Union is not liable for any use that may be made of the contained information. The resources and services used in this work were provided by the VSC (Flemish Supercomputer Center), funded by the Research Foundation - Flanders (FWO) and the Flemish Government.}$,
Thomas Dietzen$^{1}$,
Toon van Waterschoot$^{1*}$, 
}

\address{$^1$ KU Leuven, Dept. of Electrical Engineering (ESAT-STADIUS), Leuven, Belgium, \\
\{stefano.damiano, thomas.dietzen, toon.vanwaterschoot\}@esat.kuleuven.be\\
}

\begin{document}

\ninept
\maketitle

\begin{sloppy}

\begin{abstract}
The identification of siren sounds in urban soundscapes is a crucial safety aspect for smart vehicles and has been widely addressed by means of neural networks that ensure robustness to both the diversity of siren signals and the strong and unstructured background noise characterizing traffic.
Convolutional neural networks analyzing spectrogram features of incoming signals achieve state-of-the-art performance when enough training data capturing the diversity of the target acoustic scenes is available. In practice, data is usually limited and algorithms should be robust to adapt to unseen acoustic conditions without requiring extensive datasets for re-training. 
In this work, given the harmonic nature of siren signals, characterized by a periodically evolving fundamental frequency, we propose a low-complexity feature extraction method based on frequency tracking using a single-parameter adaptive notch filter. The features are then used to design a small-scale convolutional network suitable for training with limited data. 
The evaluation results indicate that the proposed model consistently outperforms the traditional spectrogram-based model when limited training data is available, achieves better \mbox{cross-domain generalization and has a smaller size.}
\end{abstract}

\begin{keywords}
siren detection, frequency tracking, data-efficient learning, convolutional neural network
\end{keywords}

\section{Introduction}
\label{sec:introduction}

The increasing level of automation of road vehicles requires robust systems that enable cars to understand their surroundings and either provide feedback to human drivers or autonomously interact with other road users. Environmental awareness is obtained by collecting information using multi-modal sensors including cameras, radar, lidar and acoustic sensors~\cite{marchegiani_how_2022}. With the rich urban soundscape containing information on events happening on a road, sound detection has been widely explored for both monitoring purposes~\cite{wonIntelligentTrafficMonitoring2020,damiano_can_2024} and to identify emergency or harmful situations that require attention~\cite{walden_improving_2022, yin_real-time_2023,furletov_auditory_2021, marchegiani_listening_2022,nandwanaSmartCarsThatCan2016, leeEnsembleConvolutionalNeural2017, ramirezSirenIdentificationSystem2022, pramanickDeepLearningBased2021}. In particular, emergency vehicles (EV) are usually announced by the sound of their siren that can often be detected from a distance, before they become visible to the driver or can be identified using other sensing modalities (e.g., when obstacles occlude the line of sight or the EV is behind a corner).

Several siren identification algorithms have been proposed, with deep learning models achieving state-of-the-art performance thanks to their robustness to the diversity of siren signals (three classes of sirens exist, namely two-tone, wail and yelp, and a large variability can be observed even between sirens of the same type) and to the prominent and non-stationary traffic background noise~\cite{walden_improving_2022, marchegiani_listening_2022, tranAcousticBasedEmergencyVehicle2020, cantariniAcousticFeaturesDeep2021, cantariniFewShotEmergencySiren2022, damianoSyntheticDataGeneration2024}. Most state-of-the-art solutions rely on a spectrogram-based time-frequency representation of sound signals fed to 2D convolutional neural networks (CNN)~\cite{cantariniAcousticFeaturesDeep2021, tranAcousticBasedEmergencyVehicle2020, damianoSyntheticDataGeneration2024}. These vision-inspired architectures process the spectrogram as a 2D image and achieve high accuracy when (diverse) enough training data is at disposal.
Siren identification systems are faced with several use-case specific challenges. First, models to be deployed on-vehicle should have a low complexity to run on resource-constrained embedded devices. Second, models should have a vast generalization ability to face the diverse urban soundscape: not only the background noise can significantly differ based on factors such as the landscape (e.g., urban vs. rural), the region, or the time of the day (and day of the year), but also the characteristics of siren sounds can strongly vary among different countries. 
Finally, in practice, the amount of available data can be limited and datasets are unlikely to capture the diversity of the target scenes. In~\cite{damianoSyntheticDataGeneration2024}, the generalization ability of state-of-the-art siren identification networks is investigated, showing that models trained on one dataset do not always generalize well to unseen domains (cross-dataset setting): using synthetic data for training purposes is thus proposed to enhance data diversity. In~\cite{cantariniFewShotEmergencySiren2022}, instead, data-efficient learning is achieved by fine-tuning a pre-trained environmental audio classification model in a \emph{few-shot} setting to identify a specific type of two-tone siren.

Aiming for data-efficiency and low complexity, in this work, we propose novel features for siren identification based on frequency tracking. In contrast to the unstructured nature of traffic noise, sirens are artificial signals generated with a simple process: all types of sirens have a harmonic behavior characterized by a periodically evolving fundamental frequency, that can be tracked over time by means of an adaptive notch filter (ANF)~\cite{raoAdaptiveNotchFiltering1984, aliFrequencyTrackerBased2023}. 
Adopting the single-parameter ANF design proposed in~\cite{aliFrequencyTrackerBased2023} (KalmANF), we design a CNN model using two features, namely the tracked fundamental frequency and the power ratio between the tracked sinusoidal component, extracted by the ANF, and the full audio signal. This allows to drastically reduce the input feature size compared to using the full spectrogram, and to thus adopt low-complexity networks.
In the experimental evaluation, we show that the proposed model is suitable for training with a limited amount of data, consistently outperforming a spectrogram-based CNN~\cite{cantariniAcousticFeaturesDeep2021} when small training sets are used. Moreover, the proposed model is 7 times smaller than the baseline~\cite{cantariniAcousticFeaturesDeep2021} and achieves improved performance in a cross-dataset setting. Accompanying code is available at~\cite{damianoANFsirenDetectionCode2024}.
\begin{figure*}[t]
    \centering
    \resizebox{0.97\textwidth}{!}{
    \includegraphics[width=\textwidth]{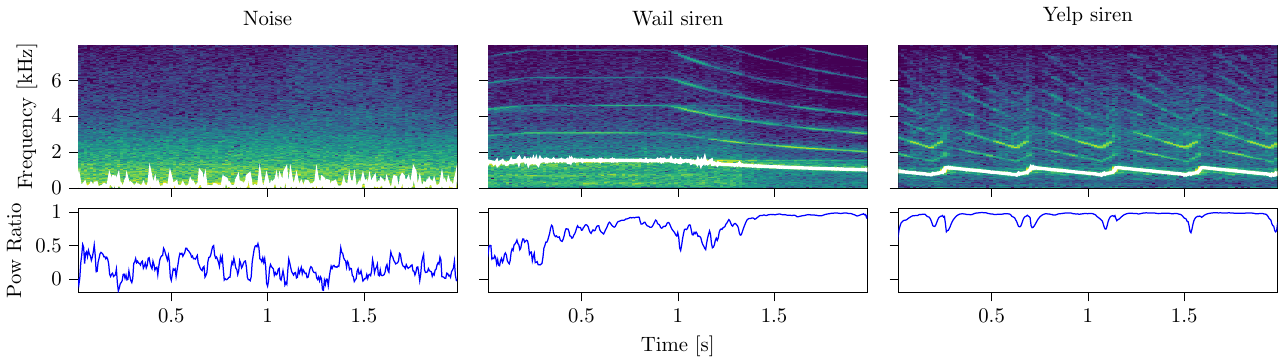}
    }
    \caption{Proposed features for three audio samples: frequency tracked by the ANF algorithm (above, highlighted in white and overlaid to the full spectrogram) and power ratio (below).}
    \label{fig:features}
\end{figure*}

\section{Problem Statement and Baseline}
\label{sec:problem_statement}
We cast siren identification as a binary classification problem, where the goal is to assign a unique label (\emph{siren} or \emph{noise}) to a $\SI{2}{\second}$ audio segment.
The task is solved using the proposed architecture (ANFNet), introduced in Sec.~\ref{sec:proposed_method}, that we compare with the spectrogram-based baseline~\cite{cantariniAcousticFeaturesDeep2021}, denoted as VGGSiren. The network is a VGG-inspired~\cite{simonyanVeryDeepConvolutional2014} 2D-CNN composed of three blocks, each containing two 2D convolutional layers and a max pooling operation, followed by a 10-neurons FC layer and the single-neuron output layer. The network takes as input the mel-spectrogram of a $\SI{2}{\second}$-long single-channel audio segment.

\begin{table}[t]
    \centering
    \begin{tabular}{ccc}
         \toprule
         Layer & Kernel Size & Filters/Neurons \\
         \midrule
         Conv1D, stride 2 &  16 & 10 \\
         MaxPool 2x1 & - & - \\
         Conv1D, stride 2 &  8 & 20 \\
         MaxPool 2x1 & - & - \\
         Conv1D, stride 2 &  4 & 40 \\
         GlobAvgPool & - & - \\
         Fully Connected & - & 40 \\
         Fully Connected & - & 20 \\
         Output & - & 1 \\
         \bottomrule
    \end{tabular}
    \caption{Proposed ANFNet architecture, taking as input the two frequency tracking features.}
    \label{tab:anfnet_architecture}
\end{table}
\section{Proposed Method}
\label{sec:proposed_method}
In this section, we summarize the KalmANF frequency tracking algorithm described in~\cite{aliFrequencyTrackerBased2023}, underlining the modifications introduced to obtain the proposed features; we then present the proposed ANFNet siren identification network.

An ANF is a type of notch filter~\cite{hiranoDesignDigitalNotch1974} whose notch frequency is recursively updated in order to suppress a high-energy sinusoidal component while leaving nearby frequencies relatively unaffected.  
The KalmanANF in \cite{aliFrequencyTrackerBased2023} is expressed
as a time-varying single-parameter bi-quadratic infinite impulse response (IIR) filter
\begin{equation}
    H(q^{-1},\,n) = \frac{1-a(n)q^{-1}+q^{-2}}{1-\rho a(n)q^{-1}+\rho^2q^{-2}}\,,
    \label{eq:biquad_anf}
\end{equation}
where $n$ is the time index, $q$  denotes the discrete-time shift operator defined such that, for an input signal $y(n)$, $q^{-k}y(n) = y(n-k)$~\cite{vanWaterschootFiftyYearsAcoustic2011}; $\rho < 1$ is a fixed hyperparameter denoting the radius of the complex conjugate pole pair and $a(n)=2\cos{[2\pi f(n)/f_s]}$ is the single filter parameter, $f(n)$ being the notch frequency and $f_s$ the sampling frequency. In the KalmANF, the time-varying coefficient $a(n)$ represents the state that is adaptively estimated in order to track the variations of $f(n)$ over time, as outlined in the following.
The given $N$-samples long input signal $y(n)$ is filtered by the direct-form II~\cite{travassos-romanoFastLeastSquares1988} of $H(q^{-1},\,n-1)$ using a joint delay line for the feedforward and the feedback paths of the IIR filter. The delay line signal $s(n)$ is defined as \cite{aliFrequencyTrackerBased2023} 
\begin{equation}
    s(n) = y(n) + \rho {a}(n-1)s(n-1) - \rho^2s(n-2)\,.
    \label{eq:measurment}
\end{equation}
In the KalmANF, $s(n)$ represents the measurement. 
This results in the state space model (see \cite{aliFrequencyTrackerBased2023} for the detailed derivation)
\begin{equation}
    \begin{bmatrix}
   a(n) \\
   1
   \end{bmatrix} 
   = 
   \begin{bmatrix}
   1 & 0 \\
   0 & 1
   \end{bmatrix} 
   \begin{bmatrix}
   a(n - 1) \\
   1
   \end{bmatrix} 
   +
   \begin{bmatrix}
   w(n) \\
   0
   \end{bmatrix} 
   \label{eq:kalman_state_update}
\end{equation}
\begin{equation}
   s(n) = 
   \begin{bmatrix}
   s(n-1) & -s(n-2) \\
   \end{bmatrix} 
   \begin{bmatrix}
   a(n) \\
   1
   \end{bmatrix} 
   +
   e(n) \,,
   \label{eq:kalman_measurement_update}
\end{equation}
where $e(n)$ is the residual signal obtained at the output of the notch filter and  $w(n)$ is the process noise.
Based on this state-space model, $a(n)$ can be estimated in a recursive manner using the Kalman filter~\cite{kalmanNewApproachLinear1960}.
The estimation procedure consists in the recursive update of the covariance of the prediction error $\hat{p}(n)$, the Kalman gain $k(n)$, and the parameter estimate $\hat{a}(n)$. These steps involve scalar operations and require a memory of 2 past samples. The filter relies on tuning three hyperparameters, namely the pole radius $\rho$, the variance $\sigma_e$ of the residual $e(n)$ and the variance $\sigma_w$ of the process noise $w(n)$. 
First, given the previous estimate $\hat{a}(n-1)$, the measurement $s(n)$ is computed according to \eqref{eq:measurment}.
Then, the covariance of the prediction error is computed as 
\begin{equation}
    \hat{p}(n|n-1) = \hat{p}(n-1) + \sigma_w
\end{equation}
and is used to obtain the Kalman gain 
\begin{equation}
    k(n) = \frac{s(n-1)}{s^2(n-1) + \frac{\sigma_e}{\hat{p}(n|n-1)}}\,.
\end{equation}
The update equation to estimate the current value of the parameter $\hat{a}(n)$ can then be expressed as
\begin{equation}
    \hat{a}(n) = \hat{a}(n-1) + k(n)e(n)\,,
\end{equation}
where, from eq.~\eqref{eq:kalman_measurement_update}, the residual takes the value 
\begin{equation}
    e(n) = s(n) - \hat{a}(n-1)s(n-1) + s(n-2)\,.
\end{equation}
Finally, the covariance of the prediction error is updated by
\begin{equation}
    \hat{p}(n) = \left( 1 - \frac{s^2(n-1)}{s^2(n-1) + \frac{\sigma_e}{\hat{p}(n|n-1)}}\right)\hat{p}(n|n-1)\,.
\end{equation} 
At each time step, the estimated parameter $\hat{a}(n)$ contains information on the frequency tracked by the ANF, that is retrieved as $\hat{f}(n) = (f_s/2\pi)\arccos{[\hat{a}(n)/2]}$ and will be used as a first feature for the siren identification network. It is important to notice that the tracked frequency is not necessarily the fundamental frequency, but the one with the highest energy. This comes with the advantage that, if the fundamental of a siren is missing or hidden in the background noise, the higher harmonics could still be tracked by the ANF. 

We then expand the above formulation to introduce a second feature that we call \emph{power ratio}, expressing the ratio between the power of the suppressed sinusoidal component and of the input signal. At each time step, the power of the input signal, the notched signal (after $\hat{f}$ has been suppressed) and the suppressed frequency component can be estimated recursively as
\begin{equation}
    P_y(n) = \lambda P_y(n-1) + (1-\lambda)y^2(n)
\end{equation}
\begin{equation}
    P_e(n) = \lambda P_e(n-1) + (1-\lambda)e^2(n)\,,
\end{equation}
\begin{equation}
    P_f(n) = P_y(n) - P_e(n)\,,
\end{equation}
where $\lambda = e^{-1/(\tau f_s)}$, with $\tau$ constituting a first additional hyperparameter representing the time constant for recursive averaging. The power ratio is finally computed as
\begin{equation}
    P_\text{ratio}(n) = P_f(n) / P_y(n)\,.
\end{equation}
To reduce the feature size, we finally downsample $P_f$ and $\hat{f}$ by a factor $q_\text{down}$, the second additional hyperparameter of the proposed method.
The procedure is summarized in Algorithm~\ref{algo:kalman_updates}.
\begin{algorithm}[t]
    \caption{The modified KalmANF algorithm}
    \label{algo:kalman_updates}
        Initialize $s(0), s(1), \hat{a}(1), \hat{p}(1) = 0$\\
        Set $\sigma_e, \sigma_w, \rho, \tau, q_\text{down}$\\
        \For{$n=2$ \textup{to} $N-1$}
        {
        $\hat{p}(n|n-1) = \hat{p}(n-1) + \sigma_w$\\
        $s(n) = y(n) + \rho\hat{a}(n-1)s(n-1)-\rho^2s(n-2)$\\
        $k(n) = \frac{s(n-1)}{s^2(n-1) + \frac{\sigma_e}{\hat{p}(n|n-1)}}$\\
        $e(n) = s(n) -\hat{a}(n-1)s(n-1) + s(n-2)$\\
        $\hat{a}(n) = \hat{a}(n-1) + k(n)e(n)$\\
        $\hat{p}(n) = \left( 1 - \frac{s^2(n-1)}{s^2(n-1) + \frac{\sigma_e}{\hat{p}(n|n-1)}}\right) \hat{p}(n|n-1)$\\
        \If{$\vert\hat{a}(n)\vert > 2$}
        {
            $\hat{a}(n) = 2\textup{sgn}(\hat{a}(n))$\\
        }
        $\hat{f}(n) = (f_s / 2\pi)\arccos{[\hat{a}(n)/2]}$\\
        $P_y(n) = \lambda P_y(n-1) + (1-\lambda)y^2(n)$\\
        $P_e(n) = \lambda P_e(n-1) + (1-\lambda)e^2(n)$\\
        $P_f(n) = P_y(n) - P_e(n)$\\
        $P_\text{ratio}(n) = P_f(n) / P_y(n)$\\
        }
        Downsample $\hat{f}$ and $P_\text{ratio}$ by factor $q_\text{down}$
\end{algorithm}

In Fig.~\ref{fig:features} the $\hat{f}$ and $P_\text{ratio}$ features are shown for a noise sample and two different siren samples (wail and yelp) extracted from the sireNNet dataset~\cite{shah_sirennet-emergency_2023}, that will be used for the experimental evaluation. In the top row, the tracked $\hat{f}$ is overlaid to the full spectrogram: nevertheless, we remark that the frequency estimate is obtained directly from the time-domain signal without computing the spectrogram. The visualization shows the effectiveness of the tracking algorithm when applied to siren signals, and underlines the clear contrast between the features extracted from a (structured) siren and the (unstructured) traffic noise.

We solve the siren identification problem using the ANFNet network, that processes the $\hat{f}$ and $P_\text{ratio}$ features extracted from a single channel, $\SI{2}{\second}$-long audio sample: the two features are stacked into a 2-channel vector provided as input to the first layer. The architecture (see Tab.~\ref{tab:anfnet_architecture}) contains three 1D convolutional layers (Conv1D) with, respectively, 10, 20 and 40 filters having kernel size 16, 8 and 4. Each of the first two Conv1D layers is followed by a max pooling operation (MaxPool) for dimensionality reduction. After the third one, a global average pooling operation (GlobAvgPool) is used as interface between the convolutional part and the classification head, composed of two fully connected (FC) layers with 40 and 20 neurons, respectively, and a single neuron output layer. We use the ReLU activation function in each hidden layer and the sigmoid activation in the output layer, and introduce dropout layers with 0.25 drop probability after each FC layer to prevent overfitting. The network has $\SI{7.7}{\kilo{}}$ floating-point 32-bit parameters.

\section{Evaluation}
\label{sec:evaluation}

We run an evaluation campaign to assess the effectiveness of the proposed method: to promote reproducibility, the code is available at~\cite{damianoANFsirenDetectionCode2024}.
For training, we use the sireNNet dataset~\cite{shah_sirennet-emergency_2023}, containing a total of 421 noise and 1254 siren samples including different types of sirens. All samples have a duration of $\SI{3}{\second}$, and since half of the siren files are artificially generated for data augmentation purposes, we exclude them and use only the 627 non-augmented siren samples. We divide this dataset into training, validation and test data with ratios $[0.8, 0.1, 0.1]$. In order to perform a data-efficient evaluation, we split the training set into subsets of different size similarly to~\cite{schmid_data-efficient_2024}: in particular, we create subsets containing an increasing percentage of the full training set, with ratios $0.25\%, 0.5\%, 1\%, 2\%, 4\%, 8\%, 16\%, 32\%, 64\%$ and $100\%$ (i.e., the entire training set). 10 folds are randomly generated for each subset, in order to compute the mean and standard deviation of the results. The subsets are created such that \begin{enumerate*}[label=(\roman*)]
    \item smaller splits are subsets of larger ones;
    \item the data distribution is kept similar to that of the entire training set;
    \item overlapping folds are allowed.
\end{enumerate*}
The validation and test sets are always used without  additional splitting. To further evaluate the generalization performance in a cross-dataset setting, we also use a subset of 210 audio files randomly extracted from the dataset~\cite{asif_large-scale_2022} (that we will call LSSiren) for testing; this dataset contains siren and noise files with lengths between $\SI{3}{\second}$ and $\SI{15}{\second}$. All files of both datasets have been re-sampled to $\SI{16}{\kilo\hertz}$ and converted to mono; moreover, since we use $\SI{2}{\second}$ samples as input, we take only the first two seconds of each file of the sireNNet dataset, and divide the LSSiren files in non-overlapping $\SI{2}{\second}$ segments. Both datasets include real recordings, with background traffic noise, moving sirens and Doppler effect.

\begin{figure}
    \centering
    \resizebox{0.93\columnwidth}{!}{
    \includegraphics{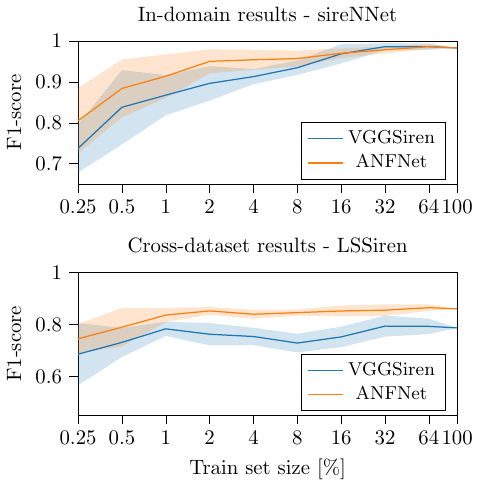}
    }
    \caption{Comparison of the average (solid line) and standard deviation (shaded area) of the F1-score for the baseline VGGSiren and the proposed ANFNet, trained with an increasing amount of data: in-domain evaluation (above) and cross-dataset evaluation (below).}
    \label{fig:results}
\end{figure}

\begin{table}[t]
    \centering
    \begin{tabular}{c|cc||cc}
        \toprule
        \multirow{2}{*}{} & \multicolumn{2}{c}{\textbf{F1-score}} & \multicolumn{2}{c}{\textbf{AUPRC}} \\
            & VGGSiren & ANFNet & VGGSiren & ANFNet \\
        \midrule
        0.25 & 0.7372 & \textbf{0.8047} & 0.8120 & \textbf{0.8471} \\
        0.5  & 0.8379 & \textbf{0.8840} & 0.8844 & \textbf{0.9162} \\
        1    & 0.8676 & \textbf{0.9139} & 0.9602 & \textbf{0.9658} \\
        2    & 0.8965 & \textbf{0.9504} & 0.9745 & \textbf{0.9787} \\
        4    & 0.9130 & \textbf{0.9543} & \textbf{0.9781} & 0.9772 \\
        8    & 0.9348 & \textbf{0.9572} & \textbf{0.9860} & 0.9796 \\
        16   & 0.9688 & \textbf{0.9702} & \textbf{0.9949} & 0.9904 \\
        32   & \textbf{0.9864} & 0.9787 & \textbf{0.9990} & 0.9962 \\
        64   & \textbf{0.9865} & \textbf{0.9865} & \textbf{0.9996} & 0.9966 \\
        100  & \textbf{0.9833} & 0.9831 & \textbf{0.9995} & \textbf{0.9995} \\
        \bottomrule
    \end{tabular}
    \caption{In-domain evaluation: average F1-score and AUPRC metrics computed on the sireNNet test set for VGGSiren and ANFNet.}
    \label{tab:metrics_in-domain}
\end{table}

\begin{table}[t]
    \centering
    \begin{tabular}{c|cc||cc}
        \toprule
        \multirow{2}{*}{} & \multicolumn{2}{c}{\textbf{F1-score}} & \multicolumn{2}{c}{\textbf{AUPRC}} \\
            & VGGSiren & ANFNet & VGGSiren & ANFNet \\
        \midrule
        0.25 & 0.6856 & \textbf{0.7448} & 0.6962 & \textbf{0.7364} \\
        0.5  & 0.7309 & \textbf{0.7895} & 0.7512 & \textbf{0.8447} \\
        1    & 0.7831 & \textbf{0.8362} & 0.8607 & \textbf{0.9169} \\
        2    & 0.7624 & \textbf{0.8522} & 0.8426 & \textbf{0.9272} \\
        4    & 0.7536 & \textbf{0.8393} & 0.8349 & \textbf{0.9147} \\
        8    & 0.7284 & \textbf{0.8455} & 0.7914 & \textbf{0.9180}  \\
        16   & 0.7520 & \textbf{0.8520} & 0.8100 & \textbf{0.9247} \\
        32   & 0.7936 & \textbf{0.8550} & 0.8492 & \textbf{0.9302} \\
        64   & 0.7926 & \textbf{0.8646} & 0.8229 & \textbf{0.9355} \\
        100  & 0.7867 & \textbf{0.8601} & 0.8052 & \textbf{0.9384} \\
        \bottomrule
    \end{tabular}
    \caption{
    Cross-dataset evaluation: average F1-score and AUPRC metrics computed on LSSiren data for VGGSiren and ANFNet.}
    \label{tab:metrics_cross-data}
\end{table}

We implement the proposed ANFNet and the baseline VGGSiren using Pytorch Lightning~\cite{Falcon_PyTorch_Lightning_2019}: for the KalmANF algorithm, we set the hyperparameters $\rho = 0.99, \sigma_w = 10^{-5}, \sigma_e = 0.66, q_\text{down} = 5, \tau = 0.02$, all chosen by manual tuning based on the best loss obtained on the validation set. For VGGSiren, to compute the mel-spectrogram we use a $1024$ samples Hann window with $512$ samples overlap, and $128$ mel channels. As a result, VGGSiren has a total of $\SI{53.9}{\kilo{}}$ floating point 32-bits parameters, thus being 7 times larger than the proposed ANFNet. For VGGSiren, we apply peak normalization to the mel-spectrograms, whereas for ANFNet we normalize the $\hat{f}$ feature to $f_s/2$ (the $P_\text{ratio}$ feature is normalized by definition). In all experiments we train both models for 400 epochs using the binary cross-entropy loss function, the Adam optimizer~\cite{kingmaAdamMethodStochastic2015} with learning rate between $0.001$ and $0.005$, a batch size between 2 and 32, both depending on the size of the training split, and select the best model based on the validation loss. To evaluate the performance, we use the F1-score~\cite{GoodBengCour16} and the area under the precision-recall curve (AUPRC)~\cite{GoodBengCour16,qi_stochastic_2024} metrics, chosen to deal with non-balanced datasets.

We train both models on the 10 folds of each sireNNet subset. Note that the $0.25\%, 0.5\%$ and $1\%$ splits contain, respectively, only $2, 4$ and $9$ samples, making the problem extremely challenging and comparable to that of few-shot learning (without pre-training). First, we evaluate in-domain performance on the sireNNet test set and report in Fig.~\ref{fig:results} the average and standard deviation (shaded area) of the F1-score. In Tab.~\ref{tab:metrics_in-domain} we report the average F1-score and AUPRC obtained with the two models for each training split. As expected, the performance of both networks degrades as the amount of training data decreases; nevertheless, ANFNet outperforms the baseline when trained using smaller subsets, and reaches a comparable performance on the larger ones (with a lower complexity).

We then evaluate the models on the LSSiren data (cross-dataset setting) and report the results in the bottom plot of Fig.~\ref{fig:results} and in Tab.~\ref{tab:metrics_cross-data}. Again, the performance of both decreases as the training dataset size decreases. In this case, the proposed ANFNet significantly outperforms the baseline on all subsets. These results indicate that the proposed features help the network capture the difference between siren and noise classes also when limited data is available, suggesting their potential for data-efficient learning. Moreover, the evaluation underlines that the proposed features ensure an enhanced robustness to domain shift compared to the mel-spectrogram. In Fig.~\ref{fig:results} it is also visible that the standard deviation is reduced compared to VGGSiren, showing that ANFNet is less sensitive to the choice of training samples. Finally, ANFNet has a lower complexity, with a 7 times smaller network size ($\SI{7.7}{\kilo{}}$ parameters vs. the $\SI{53.9}{\kilo{}}$ of VGGSiren). Note that, thanks to time-domain processing and downsampling, the feature extraction procedure has also a reduced complexity and the ANFNet input features have a smaller size compared to the mel-spectrograms used by VGGSiren. 

\section{Conclusions}
\label{sec:conclusions}
In this work, we investigated two novel features based on frequency tracking for training a siren identification model. Given the harmonic nature of siren signals, as opposed to the unstructured background noise, the features are effective for learning in a data-efficient setting, when limited data is available. The proposed system outperforms a spectrogram-based baseline on in-domain test data, when limited training data is available, and always achieves better performance in a cross-dataset setting. Moreover, its reduced complexity promotes its adoption in the automotive domain. Future work will focus on extending the frequency tracker to include higher harmonics, further investigating the generalization performance of the proposed system and optimizing the model for complexity.

\bibliographystyle{IEEEtran}
\bibliography{refs}

\begin{thebibliography}{10}
\providecommand{\url}[1]{#1}
\def\UrlFont{\rmfamily}
\providecommand{\newblock}{\relax}
\providecommand{\bibinfo}[2]{#2}
\providecommand\BIBentrySTDinterwordspacing{\spaceskip=0pt\relax}
\providecommand\BIBentryALTinterwordstretchfactor{4}
\providecommand\BIBentryALTinterwordspacing{\spaceskip=\fontdimen2\font plus
\BIBentryALTinterwordstretchfactor\fontdimen3\font minus
  \fontdimen4\font\relax}
\providecommand\BIBforeignlanguage[2]{{%
\expandafter\ifx\csname l@#1\endcsname\relax
\typeout{** WARNING: IEEEtran.bst: No hyphenation pattern has been}%
\typeout{** loaded for the language `#1'. Using the pattern for}%
\typeout{** the default language instead.}%
\else
\language=\csname l@#1\endcsname
\fi
#2}}

\bibitem{marchegiani_how_2022}
L.~Marchegiani and X.~Fafoutis, ``How {Well} {Can} {Driverless} {Vehicles}
  {Hear}? {A} {Gentle} {Introduction} to {Auditory} {Perception} for
  {Autonomous} and {Smart} {Vehicles},'' \emph{IEEE Intell. Transp. Syst.
  Mag.}, pp. 92--105, 2022.

\bibitem{wonIntelligentTrafficMonitoring2020}
M.~Won, ``Intelligent {Traffic} {Monitoring} {Systems} for {Vehicle}
  {Classification}: {A} {Survey},'' \emph{IEEE Access}, vol.~8, pp.
  73\,340--73\,358, 2020.

\bibitem{damiano_can_2024}
S.~Damiano, L.~Bondi, S.~Ghaffarzadegan, A.~Guntoro, and T.~van Waterschoot,
  ``Can synthetic data boost the training of deep acoustic vehicle counting
  networks?'' in \emph{Proc. 2024 Int. Conf. Acoust. Speech Sig. Process.
  (ICASSP)}, Seoul, South Korea, 2024, pp. 631--635.

\bibitem{walden_improving_2022}
F.~Walden, S.~Dasgupta, M.~Rahman, and M.~Islam, ``Improving the
  {Environmental} {Perception} of {Autonomous} {Vehicles} using {Deep}
  {Learning}-based {Audio} {Classification},'' \emph{arXiv:2209.04075}, Sept.
  2022.

\bibitem{yin_real-time_2023}
J.~Yin, S.~Damiano, M.~Verhelst, T.~van Waterschoot, and A.~Guntoro,
  ``Real-{Time} {Acoustic} {Perception} for {Automotive} {Applications},'' in
  \emph{2023 {Design}, {Automation} \& {Test} in {Europe} {Conf.} {Exhib.}
  ({DATE})}, Antwerp, Belgium, Apr. 2023, pp. 1--6.

\bibitem{furletov_auditory_2021}
Y.~Furletov, V.~Willert, and J.~Adamy, ``Auditory {Scene} {Understanding} for
  {Autonomous} {Driving},'' in \emph{2021 {IEEE} {Intell.} {Vehicles} {Symp.}
  ({IV})}, Nagoya, Japan, July 2021, pp. 697--702.

\bibitem{marchegiani_listening_2022}
L.~Marchegiani and P.~Newman, ``Listening for {Sirens}: {Locating} and
  {Classifying} {Acoustic} {Alarms} in {City} {Scenes},'' \emph{IEEE Trans.
  Intell. Transp. Syst.}, vol.~23, no.~10, pp. 17\,087--17\,096, 2022.

\bibitem{nandwanaSmartCarsThatCan2016}
M.~K. Nandwana and T.~Hasan, ``Towards {Smart}-{Cars} {That} {Can} {Listen}:
  {Abnormal} {Acoustic} {Event} {Detection} on the {Road},'' in \emph{Proc.
  2016 Interspeech}, San Francisco, USA, Sept. 2016, pp. 2968--2971.

\bibitem{leeEnsembleConvolutionalNeural2017}
D.~Lee, S.~Lee, Y.~Han, and K.~Lee, ``Ensemble of {Convolutional} {Neural}
  {Networks} for {Weakly}-{Supervised} {Sound} {Event} {Detection} {Using}
  {Multiple} {Scale} {Input},'' in \emph{Proc. {Detection} {Classification}
  {Acoust.} {Scenes} {Events} 2017 {Workshop} ({DCASE2017})}, Munich, Germany,
  Nov. 2017.

\bibitem{ramirezSirenIdentificationSystem2022}
A.~E. Ramirez, E.~Donati, and C.~Chousidis, ``A siren identification system
  using deep learning to aid hearing-impaired people,'' \emph{Engineering
  Applications of Artificial Intelligence}, vol. 114, p. 105000, 2022.

\bibitem{pramanickDeepLearningBased2021}
D.~Pramanick, H.~Ansar, H.~Kumar, S.~Pranav, R.~Tengshe, and B.~Fatimah, ``Deep
  learning based urban sound classification and ambulance siren detector using
  spectrogram,'' in \emph{Proc. 12th Int. Conf. Computing Comm. Networking
  Technologies (ICCCNT)}, Kharagpur, India, 2021, pp. 1--6.

\bibitem{tranAcousticBasedEmergencyVehicle2020}
V.-T. Tran and W.-H. Tsai, ``Acoustic-{Based} {Emergency} {Vehicle} {Detection}
  {Using} {Convolutional} {Neural} {Networks},'' \emph{IEEE Access}, vol.~8,
  pp. 75\,702--75\,713, 2020.

\bibitem{cantariniAcousticFeaturesDeep2021}
M.~Cantarini, A.~Brocanelli, L.~Gabrielli, and S.~Squartini, ``Acoustic
  {Features} for {Deep} {Learning}-{Based} {Models} for {Emergency} {Siren}
  {Detection}: {An} {Evaluation} {Study},'' in \emph{2021 12th {Int.} {Symp.}
  {Image} {Sig.} {Process.} {Anal.} ({ISPA})}, Zagreb, Croatia, Sept. 2021, pp.
  47--53.

\bibitem{cantariniFewShotEmergencySiren2022}
M.~Cantarini, L.~Gabrielli, and S.~Squartini, ``Few-{Shot} {Emergency} {Siren}
  {Detection},'' \emph{Sensors}, vol.~22, no.~12, p. 4338, June 2022.

\bibitem{damianoSyntheticDataGeneration2024}
S.~Damiano, B.~Cramer, A.~Guntoro, and T.~van Waterschoot, ``Synthetic {Data}
  {Generation} {Techniques} for {Training} {Deep} {Acoustic} {Siren}
  {Identification} {Networks},'' \emph{Frontiers {Sig.} {Process.}}, vol.~4,
  2024.

\bibitem{raoAdaptiveNotchFiltering1984}
D.~Rao and {Sun-Yuan Kung}, ``Adaptive notch filtering for the retrieval of
  sinusoids in noise,'' \emph{IEEE Trans. Acoust. Speech Sig. Process.},
  vol.~32, no.~4, pp. 791--802, Aug. 1984.

\bibitem{aliFrequencyTrackerBased2023}
R.~Ali and T.~van Waterschoot, ``A {Frequency} {Tracker} {Based} on a {Kalman}
  {Filter} {Update} of a {Single} {Parameter} {Adaptive} {Notch} {Filter},'' in
  \emph{Proc. 26th {Int}. {Conf}. {Digital} {Audio} {Effects} ({DAFx})},
  Copenhagen, Denmark, Sept. 2023.

\bibitem{damianoANFsirenDetectionCode2024}
\BIBentryALTinterwordspacing
S.~Damiano and T.~Dietzen, ``An {ANF}-based siren identification system,''
  {Github} Repository, 2024. [Online]. Available:
  \url{https://github.com/steDamiano/anf-siren-identification}
\BIBentrySTDinterwordspacing

\bibitem{simonyanVeryDeepConvolutional2014}
K.~Simonyan and A.~Zisserman, ``Very deep convolutional networks for
  large-scale image recognition,'' \emph{arXiv:1409.1556}, 2014.

\bibitem{hiranoDesignDigitalNotch1974}
K.~Hirano, S.~Nishimura, and S.~Mitra, ``Design of {Digital} {Notch}
  {Filters},'' \emph{IEEE Trans. Commun.}, vol.~22, no.~7, pp. 964--970, July
  1974.

\bibitem{vanWaterschootFiftyYearsAcoustic2011}
T.~van Waterschoot and M.~Moonen, ``Fifty years of acoustic feedback control:
  State of the art and future challenges,'' \emph{Proc. IEEE}, vol.~99, no.~2,
  pp. 288--327, 2011.

\bibitem{travassos-romanoFastLeastSquares1988}
J.~Travassos-Romano and M.~Bellanger, ``Fast least squares adaptive notch
  filtering,'' in \emph{Proc. {Int.} {Conf.} {Acoust.}, {Speech} {Sig.}
  {Process.} {ICASSP} 1988}, New York, NY, USA, 1988, pp. 1391--1394.

\bibitem{kalmanNewApproachLinear1960}
R.~E. Kalman, ``A new approach to linear filtering and prediction problems,''
  \emph{Journal of Basic Engineering}, vol.~82, no.~1, pp. 35--45, Mar. 1960.

\bibitem{shah_sirennet-emergency_2023}
A.~Shah and A.~Singh, ``{sireNNet}-{Emergency} {Vehicle} {Siren}
  {Classification} {Dataset} {For} {Urban} {Applications},'' Mendeley Data,
  2023, doi: 10.17632/j4ydzzv4kb.1.

\bibitem{schmid_data-efficient_2024}
F.~Schmid, P.~Primus, T.~Heittola, A.~Mesaros, I.~Martín-Morató, K.~Koutini,
  and G.~Widmer, ``Data-efficient low-complexity acoustic scene classification
  in the dcase 2024 challenge,'' \emph{arXiv:1706.10006}, 2024.

\bibitem{asif_large-scale_2022}
M.~Asif, M.~Usaid, M.~Rashid, T.~Rajab, S.~Hussain, and S.~Wasi, ``Large-scale
  audio dataset for emergency vehicle sirens and road noises,''
  \emph{Scientific Data}, vol.~9, no.~1, p. 599, Oct. 2022.

\bibitem{Falcon_PyTorch_Lightning_2019}
\BIBentryALTinterwordspacing
W.~Falcon and {The PyTorch Lightning team}, ``{PyTorch Lightning},'' {Github}
  Repository, Mar. 2019. [Online]. Available:
  \url{https://github.com/Lightning-AI/lightning}
\BIBentrySTDinterwordspacing

\bibitem{kingmaAdamMethodStochastic2015}
D.~P. Kingma and J.~Ba, ``Adam: {A} method for stochastic optimization,'' in
  \emph{Proceedings of the 3rd {International} {Conference} on {Learning}
  {Representations} ({ICLR})}, San Diego, USA, 2015.

\bibitem{GoodBengCour16}
I.~J. Goodfellow, Y.~Bengio, and A.~Courville, \emph{Deep Learning}.\hskip 1em
  plus 0.5em minus 0.4em\relax Cambridge, MA, USA: MIT Press, 2016.

\bibitem{qi_stochastic_2024}
Q.~Qi, Y.~Luo, Z.~Xu, S.~Ji, and T.~Yang, ``Stochastic optimization of areas
  under precision-recall curves with provable convergence,'' in \emph{Proc.
  35th Int. Conf. Neural Information Process. Syst.}, online, 2021.

\end{thebibliography}

\end{sloppy}
\end{document}